\documentclass[aip,reprint]{revtex4-1}
\usepackage{epsfig,amssymb,amsmath}

\draft 

\begin{document}


\title{Periodicity of magnetization reversals in $\varphi_0$ Josephson junction} 



\author{P.~Kh.~Atanasova}
\affiliation{University of Plovdiv Paisii Hilendarski, 24 Tzar Asen, 4000 Plovdiv, Bulgaria}

\author{S.~A.~Panayotova}
\affiliation{University of Plovdiv Paisii Hilendarski, 24 Tzar Asen, 4000 Plovdiv, Bulgaria}

\author{I.~R.~Rahmonov}
\affiliation{Joint Institute of Nuclear Research, Dubna, Moscow Region, 141980, Russia}
\affiliation{Umarov Physical and Technical Institute, TAS, Dushanbe, 734063, Tajikistan}

\author{Yu. M. Shukrinov}
\affiliation{Joint Institute of Nuclear Research, Dubna, Moscow Region, 141980, Russia}
\affiliation{Dubna State University, Dubna,  141980, Moscow Region, Russia}

\author{E.~V.~Zemlyanaya}
\affiliation{Joint Institute of Nuclear Research, Dubna, Moscow Region, 141980, Russia}
\affiliation{Dubna State University, Dubna,  141980, Moscow Region, Russia}

\author{M.~V.~Bashashin}
\affiliation{Joint Institute of Nuclear Research, Dubna, Moscow Region, 141980, Russia}
\affiliation{Dubna State University, Dubna,  141980, Moscow Region, Russia}

\date{\today}

\begin{abstract}
The magnetization reversal in ${\varphi_0}$-Josephson junction with direct coupling between magnetic moment and Josephson current has been studied. By adding pulse signal, the dynamics of magnetic moment components have been simulated and the full magnetization reversal at different parameters of the junction has been demonstrated. We obtain a detailed pictures representing the intervals of the damping parameter $\alpha$, Josephson to magnetic energy relation $G$ and spin-orbit coupling parameter $r$ with full magnetization reversal. A periodicity in the appearance of magnetization reversal intervals with increase in Josephson to magnetic energy relation is found. The obtained  results might be used in different fields of superconducting spintronics.
\end{abstract}

\maketitle
The superconducting spintronics based on the interaction of superconducting current with magnetic moment in Josephson superconductor-ferromagnet structures attracts much attention today due to possibility of controlling magnetism by superconductivity and due to a perspective of applications in quantum computer technologies~\cite{golubov17,zhu17,houzet08,petkovic09,jj15,cai10,konschele15,chud16}. Magnetization reversal (MR) by superconducting current based on the fact that the equilibrium orientation of the magnetic moment of a ferromagnet is determined by the magnetic anisotropy that is of relativistic spin-orbit origin and is comparable to the superconducting gap \cite{chud17}. In Ref.\cite{chud17} the authors have demonstrated that 2D superconductors with
large spin-orbit coupling present  an opportunity to manipulate a nanoscale magnetic moment embedded in a
single superconducting layer by a superconducting current.  They showed that a transport current through the superconductor with spin-orbit coupling generates an effective magnetic field in the cluster that is capable of switching the direction of the magnetic moment between two opposite equilibrium orientations along the easy anisotropy axis. It may lead to a significant rate of quantum tunneling of the magnetic moment, providing a possible design for a qubit.

In the superconductor-ferromagnetic-superconductor (SFS) structures, the spin-orbit coupling in ferromagnetic layer without inversion symmetry provides a mechanism for a direct (linear) coupling between the magnetic moment and the superconducting current\cite{buzdin-prl08}. Such Josephson junctions are called $\varphi_{0}$-junction. The possibility of controlling the magnetic properties by means of the superconducting current, and as well the effect of magnetic dynamics on the superconducting current attracts an intensive attention today\cite{buzdin-prl08, buzdin-rmp05, Konschelle-prb09, srsb17, borovets2018}.
In Ref.\cite{srsb17} a realization of MR was demonstrated using the numerical simulation of its dynamics. It was shown that MR is very sensitive to the model parameters. Until now effect of system parameters was not investigated in detailed. Due to complexity of the system under study, the question concerning prediction if  the full reversal would be realized at fixed parameters of JJ and current pulse is still open.

In this paper we investigate the effect of model parameters on the full MR in the ${\varphi_0}$-Josephson junction. We have established that the realization of MR is characterized by some periodicity on damping parameter, relation of Josephson to magnetic energy and spin-orbit coupling parameter. The obtained results might be useful for understanding of complex physical processes in different fields of superconductor spintronics.

\begin{figure}[h!]
\centering\includegraphics[width=8.cm]{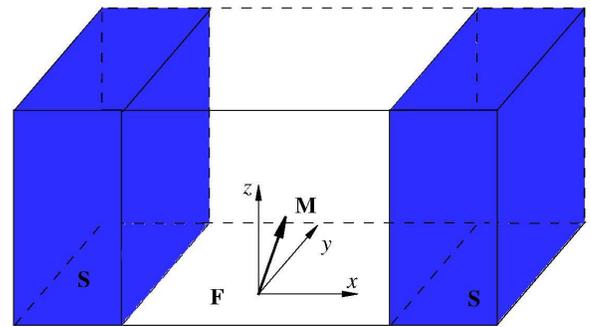}
{\caption {Geometry of the considered SFS Josephson junction. In this scheme S indicates the superconducting layers and F indicates the ferromagnetic one. Here $\textbf{M}$ is the magnetization vector and $z$ direction is it's easy axis. The external current pulse flows to the $x$ direction.}
\label{zad2_fiz}}
\end{figure}

Geometry of the superconductor-ferromagnetic-superconductor Josephson junction (SFS JJ) under consideration is presented in Fig.~\ref{zad2_fiz}. The ferromagnetic easy-axis is directed along the $z$-axis, which is also the direction of gradient of the spin-orbit potential. The magnetization component $m_y$ is coupled with Josephson current $I_s$, which is along the $x$-axis.

The magnetization dynamics of our system is described by the
Landau-Lifshitz-Gilbert equation \cite{srsb17} where effective field depends on phase difference
\begin{eqnarray}
\frac{d {\bf M}}{dt} &=& -\gamma {\bf M} \times {\bf
H_{eff}}+\frac{\alpha}{M_{0}}\bigg({\bf M}\times \frac{d {\bf
M}}{dt} \bigg) \nonumber\\
{\bf H_{eff}} &=& \frac{K}{M_{0}}\bigg[G r \sin\bigg(\varphi - r
\frac{M_{y}}{M_{0}} \bigg) {\bf\widehat{y}} +
\frac{M_{z}}{M_{0}}{\bf\widehat{z}}\bigg], \label{llg1}
\end{eqnarray}
where $\gamma$ is the gyromagnetic ratio, $\alpha$ is a
phenomenological damping constant, $M_{0}=\|{\bf M}\|$,
$\displaystyle G= E_{J}/(K \mathcal{V})$--relation of Josephson energy to the anisotropic one, $K$ is an anisotropic constant, $\mathcal{V}$ is the volume of ferromagnetic layer,  $l=4
h L/\hbar \upsilon_{F}$, $L$ is the length of $F$ layer, and $h$
denotes the exchange field in the ferromagnetic layer, $r$ is spin--orbit coupling parameter.

The system of equations in dimensionless form can be written as:

 \begin{eqnarray}\label{system4odu}
\vspace{0.3cm}
\nonumber
\displaystyle\frac{dm_x}{dt} &=& -\frac{1}{1+M\alpha^{2}}\{(m_yH_z - m_zH_y) \nonumber\\
&+& \alpha[m_x( m_xH_x + m_y H_y + m_zH_z) - H_x]\}\nonumber\\
\vspace{0.3cm}
\displaystyle\frac{dm_y}{dt} &=& -\frac{1}{1+M\alpha^{2}}\{(m_zH_x - m_xH_z)\\
&+& \alpha[m_y( m_xH_x + m_y H_y + m_zH_z) - H_y]\}\nonumber\\
\vspace{0.3cm}
\nonumber
\displaystyle\frac{dm_z}{dt} &=& -\frac{1}{1+M\alpha^{2}}\{(m_xH_y - m_yH_x) \nonumber\\
&+& \alpha[m_z( m_xH_x + m_y H_y + m_zH_z) - H_z]\}\nonumber
\end{eqnarray}
where $m_{i}$ are the components of magnetization and $H_i$, are components of the effective magnetic field, which are given by the expressions

 \begin{eqnarray}
 \label{h_eff}
H_x(t) &=& 0,\nonumber\\
H_y(t) &=& G r\sin(\varphi(t)-r m_y(t))\\
H_z(t) &=& m_z(t)\nonumber
\end{eqnarray}

In the system of equations (\ref{system4odu}), the time is normalized to the $\omega_{F}^{-1}$ (where $\omega_{F}$ is feromagnetic resource frequency), $m_{i}$ is norm to the $M_{0}$.
In order to solve system of equations taking into account expressions (\ref{h_eff}), we need to clarify the phase difference $\varphi$. The equation for phase difference can be written using Resistively Capacitively Shunted Junction (RCSJ)--model~\cite{likharev}. For simplicity, here we consider the JJ with low capacitance C ( $R^{2}C/L_{J}<<1$, where $L_{J}$ is the inductance of the JJ and $R$ is its resistance), i.e., we do not take into account the displacement current. In this case the electric current through JJs is
\begin{equation}
I_{pulse}= w\frac{d\varphi}{dt}+\sin(\varphi-rm_{y})
\label{current}
\end{equation}
where $w=\frac{V_F}{I_cR}=\frac{\omega_F}{\omega_R}$, $V_F=\frac{\hbar\omega_F}{2e}$, $I_c$ - critical current, $R$- resistance of JJ, $\omega_R =\frac{2e I_cR}{\hbar}$ - characteristic Josephson frequency at $I=I_{c}$.

Initial conditions for these time-dependent functions are:
\begin{equation}\label{initialcond}
 m_x(0) = 0,\quad m_y(0) = 0,\quad m_z(0) = 1,\quad \varphi(0) = 0.
\end{equation}

The simulations have been done with the user software \cite{mmcp2017} complemented by the implicit two-stage Gauss-Legendre method, providing a higher accuracy than the explicit Runge-Kutta scheme, for numerical solution of the the system (\ref{system4odu}) with initial conditions~\cite{borovets2018} (\ref{initialcond}). We investigated an influence of the model parameters and of the current pulse shape on the MR in $\varphi_0$ Josephson junction\cite{srsb17,shukrinov_ieee-2018,cai10}.

To demonstrate the MR we have used rectangular form current pulse.

\begin{eqnarray}
\label{gauss1}
I_{pulse}(t) = \left\{
                 \begin{array}{ll}
                   A_s, & t \in [t_0 - 1/2 \Delta t, t_0 + 1/2 \Delta t];\\
                  0, & \textrm{otherwise},
                 \end{array}
               \right.
\end{eqnarray}
with the amplitude $A_s$ and duration $\Delta t$. We solve numerically the system of equations (\ref{system4odu}) together with equation (\ref{current}) using (\ref{gauss1}).

\begin{figure}[h!]
\centering
\includegraphics[width=7cm]{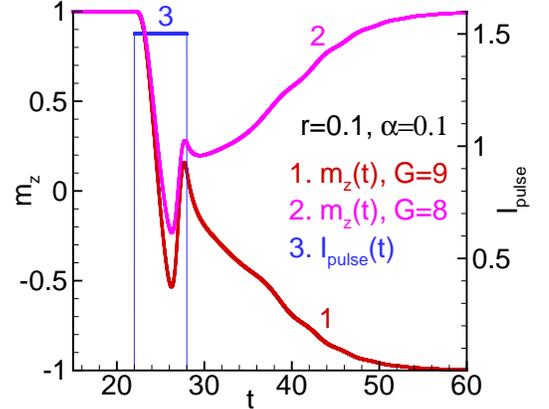}
\caption {Demonstration of the effect of Josephson $\varphi_0$ magnetic energy relation on the magnetization reversal.}
\label{reversal}
\end{figure}

First, we demonstrate the examples of MR which can be seen in Fig.~\ref{reversal}, where the time dependencies of $m_{z}$ for $G=9$ (line 1), and $G=8$ (line 2) and applied external current pulse (line 3) are presented. The calculations are performed for the spin-orbit coupling parameter $r=0.1$, dissipation $\alpha=0.1$, signal amplitude $A_{s}=1.5$. As we can see, the realization of MR strongly depends on the model parameters. Therefore, the determination of parameter intervals where MR can be realized is a very important problem to clarify the features of this phenomena.

\begin{figure}[h!]
\centering\includegraphics[width=7.cm]{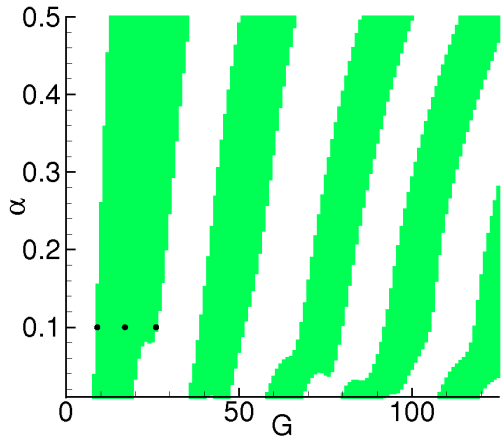}
\caption {The $G-\alpha$ diagram of the MR realization. The results are obtained with G-step $\Delta G=1$ and $\alpha$-step $\Delta \alpha=0.001$ at $A_s=1.5$, $r = 0.1$, $t_{0} = 25$,
$\Delta t = 6$, $\omega_F = 1$, $h=0.01$. The dots mark the $G$ and $r$ values, at which time dependence of $m_{z}$ is demonstrated in Fig.\ref{g-dep}.}
\label{r0.1_alpha0.6_G200}
\end{figure}

In order to define the intervals of G for the different values of damping parameter $\alpha$,
 where the MR can be realized, we have calculated the time dependencies for the different values of $G$ and $\alpha$. Then we have selected values of parameters where the MR occurs. The calculations are performed for the values of $G$ from $G=1$ up to $G=130$ with the stepsize $\Delta G=1$ and for the values of $\alpha$ from $\alpha=0.01$ up to $\alpha=0.5$ with the stepsize $\Delta \alpha=0.001$.
  For every couple of values $(\alpha, G)$ the system (\ref{system4odu}),(\ref{initialcond}) is solved by the Gauss-Legendre method with the time step $h = 0.01$ in the interval $t\in[0,T_{max}]$, $T_{max}=200$. This method provides the fourth accuracy order that means $O(h^4) \approx 10^{-8}$. At the end time $t=T_{max}$, the inequality $|m_z + 1| \leq 0.0001$  was checked to identify if the MR occured. If so, the respective values of $\alpha$ and $G$ are appended to the vector of the MR points and are saved in outer data file. These results are visualized in Fig.~\ref{r0.1_alpha0.6_G200}.

The value of the spin-orbit coupling parameter is taken to be $r = 0.1$.
Here we also observe that the effect of MR is not observed at the beginning of $G$. After that it is periodically seen for every layer on $\alpha$.

We found some periodic dependence in the appearance of MR intervals with increase in $G$. At small $G$ the width of these intervals is increasing with increase in $\alpha$.

\begin{figure}[h!]
\centering
\includegraphics[width=3.5cm]{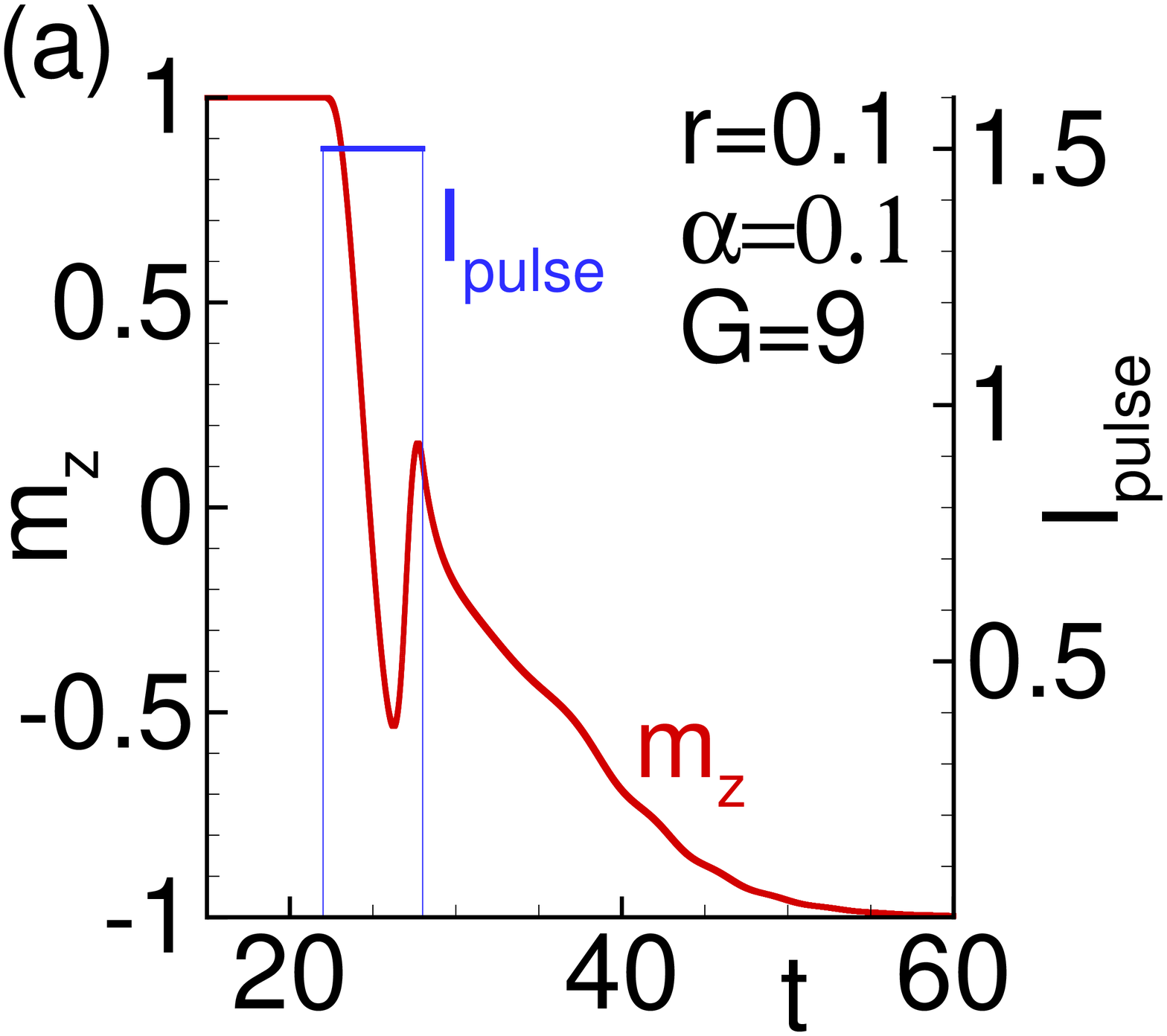}\hspace{0.2cm}
\includegraphics[width=3.5cm]{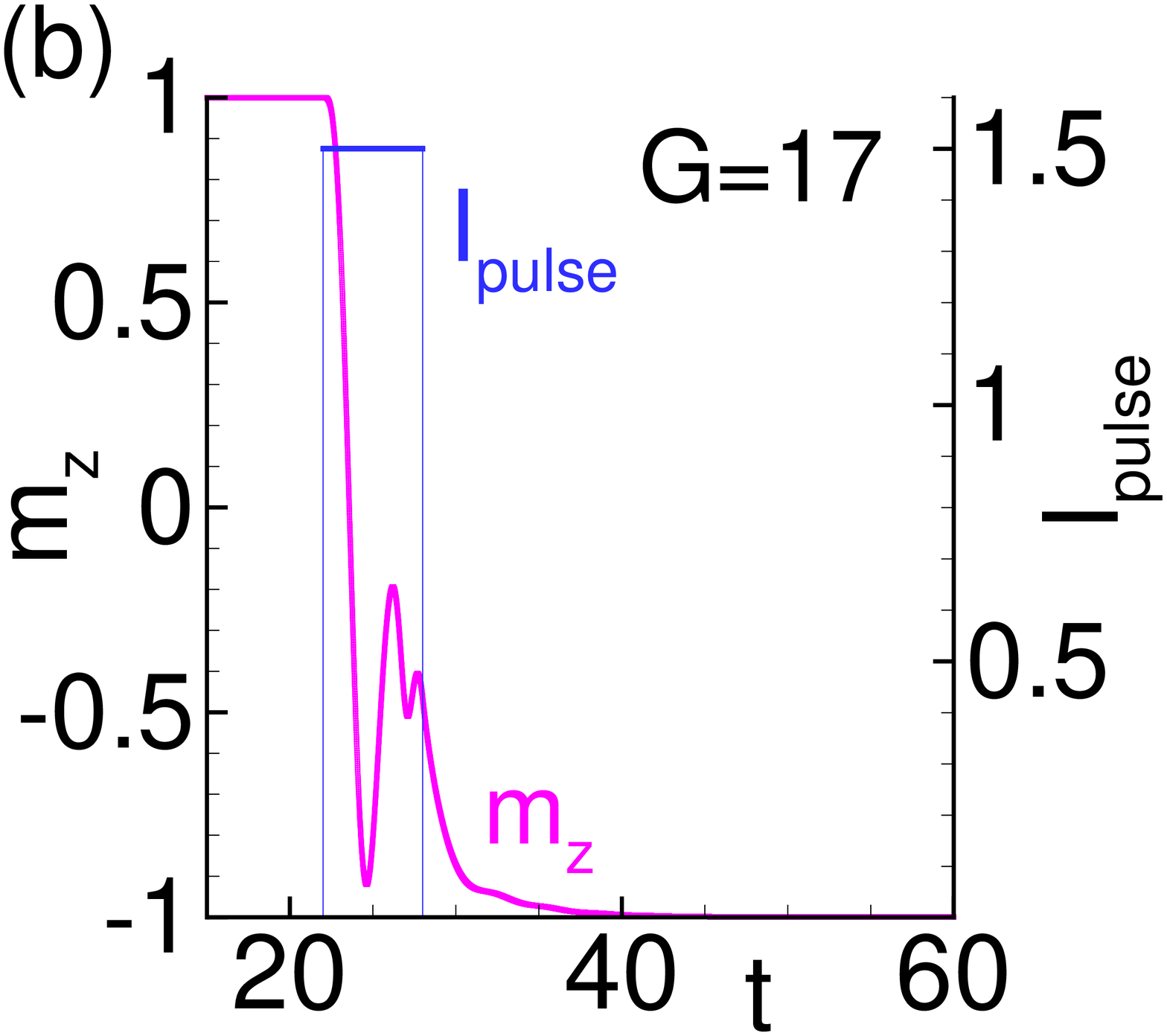}\\
\includegraphics[width=3.5cm]{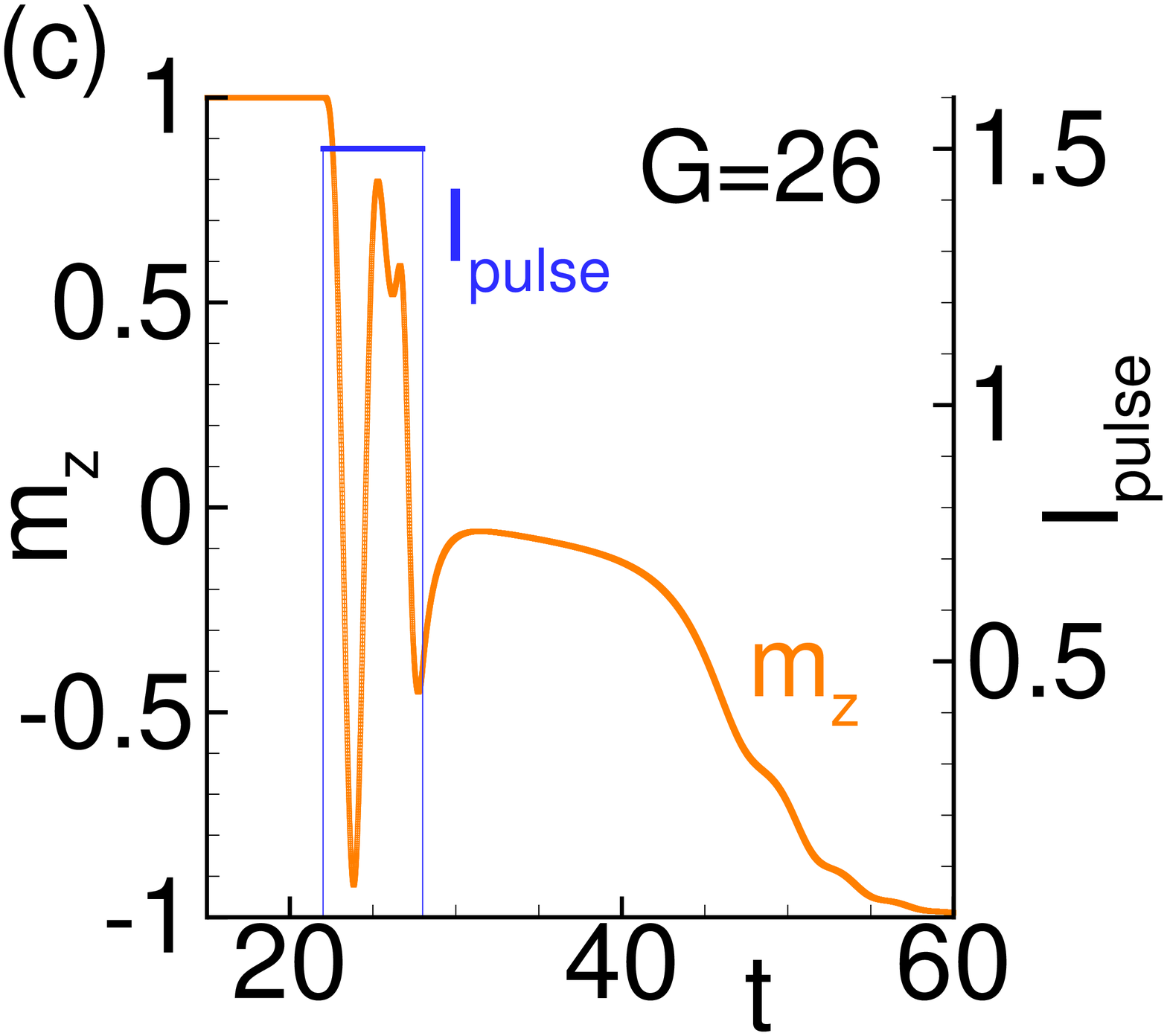}
\caption {(a)Time dependence of the magnetization component $m_{z}$ for $G=9$; (b) The same as in the case (a) for $G=17$; (c) The same as in the case (a) for $G=26$. }
\label{g-dep}
\end{figure}

The dynamics of $m_{z}$ inside the obtained interval of $G$ is presented in Fig.\ref{g-dep}. The calculations are performed for the spin-orbit coupling parameter $r=0.1$, dissipation $\alpha=0.1$, signal amplitude $A_{s}=1.5$ at the different values of energy relation $G=9$, $G=17$, and $G=26$. As we see, with an increase in $G$, the number of oscillations of $m_{z}$ is increased inside the time interval corresponding to the current pulse. The other parameters were $A_s=1.5, t_0 = 25, \Delta t = 6, \omega_F = 1, w=1$.

\begin{figure}[h!]
\centering\includegraphics[width=7.cm]{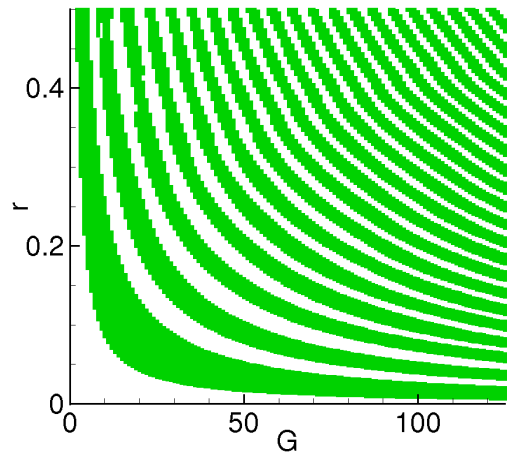}
\caption {Demonstration of intervals of complete MR in $r-G$-plane. The results are obtained with G-step $\Delta G=1$ and $r$-step $\Delta r=0.01$ at $A_s=1.5; \alpha = 0.5;  t_0 = 25; \Delta t = 6;\omega_F = 1$, $h=0.01$.}
\label{alpha0.5_r_1_G100}
\end{figure}

Results of the MR simulation on the $r-G$-plane are presented in  Fig.~\ref{alpha0.5_r_1_G100}. An increase in spin-orbit coupling leads to the shifting of the MR domains to the region of small G, with decreasing of their width.

As summary, the full MR in ${\varphi_0}$-Josephson junction with direct coupling between magnetic moment and Josephson current at different parameters of the junction and external signal has been analysed depending on parameters of the model. We obtained the detailed pictures representing the charts of the damping parameter, Josephson to magnetic energy relation and the spin-orbit coupling parameter where the full MR occurs. A periodic dependence in the appearance of reversal intervals with increase in Josephson to magnetic energy relation is observed. One expects that the obtained results might be useful in different fields of superconducting spintronics.

The authors thank K.~Sengupta for helpful discussions. The investigations are supported by the project FP17-FMI-008, Bulgaria, by the JINR-Bulgaria Cooperation Program and by the RFBR in the framework of research grants 18-02-00318, 18-52-45011. The numerical simulations are supported by the RSF in the framework of project 18-71-10095.

\end{document}